\documentclass[a4paper, 11pt]{article}
\usepackage{jinstpub} 
\usepackage{textcomp} 
\usepackage{changes}
\usepackage{booktabs}
\usepackage{multirow}
\usepackage{multicol}
\usepackage[noabbrev]{cleveref} 
\usepackage{xspace}
\usepackage{array}
\usepackage{longtable}
\usepackage{lineno}
\usepackage[super]{nth} 

\usepackage[normalem]{ulem} 

\usepackage{listings}
\definecolor{backcolour}{rgb}{0.95,0.95,0.92}
\lstdefinestyle{ThisStyle}{
    backgroundcolor=\color{backcolour},
    basicstyle=\ttfamily\footnotesize,
    breakatwhitespace=false,
    breaklines=true,
    numbersep=5pt,
    showspaces=false,
    showtabs=false,
    tabsize=2
}
\crefname{equation}{eq.}{eqs.}           
\Crefname{equation}{Equation}{Equations} 

\begin{document}

\nolinenumbers

\title{Operation of the Trigger System for the ICARUS Detector at Fermilab}

\collaboration{ICARUS collaboration}

%
%
%
%
%
%
%
%
%

\author[a]{F.~Abd~Alrahman,}
\author[b]{P.~Abratenko,}
\author[c]{N.~Abrego-Martinez,}
\author[a]{A.~Aduszkiewicz,}
\author[d]{F.~Akbar,}
\author[e]{L.~Aliaga~Soplin,}
\author[f]{M.~Artero~Pons,}
\author[e]{J.~Asaadi,}
\author[g]{W.~F.~Badgett,}
\author[f]{B.~Baibussinov,}
% ...

%
%

\affiliation[a]{University of Houston, Houston, TX 77204, USA}
\affiliation[b]{Tufts University, Medford, MA 02155, USA}
\affiliation[c]{Centro de Investigacion y de Estudios Avanzados del IPN (Cinvestav), Mexico City}
\affiliation[d]{University of Rochester, Rochester, NY 14627, USA}
\affiliation[e]{University of Texas at Arlington, Arlington, TX 76019, USA}
\affiliation[f]{INFN Sezione di Padova and University, Padova, Italy}
\affiliation[g]{Fermi National Accelerator Laboratory, Batavia, IL 60510, USA}
% ...

\emailAdd{petrillo@slac.stanford.edu}

\date{\today}

\newcommand{\RunOne}{Run1}
\newcommand{\RunTwo}{Run2}
\newcommand{\RunThree}{Run3}

\newcommand{\Order}[1]{\ensuremath{\mathcal{O}\left(#1\right)}\xspace}

\newcommand{\dEdx}{\ensuremath{\text{d}E/\text{d}x}\xspace}

\newcommand{\dCRT}{\ensuremath{d_{\text{CRT}}}}

\newcommand{\measValue}[2]{\ensuremath{#1}\,\text{#2}}

\newcommand{\ps}[1]{\measValue{#1}{ps}}
\newcommand{\ns}[1]{\measValue{#1}{ns}}
\newcommand{\us}[1]{\measValue{#1}{\textmu{}s}}
\newcommand{\ms}[1]{\measValue{#1}{ms}}

\newcommand{\m}[1]{\measValue{#1}{m}}
\newcommand{\cm}[1]{\measValue{#1}{cm}}
\newcommand{\mm}[1]{\measValue{#1}{mm}}

\newcommand{\sqm}[1]{\measValue{#1}{\ensuremath{\text{m}^{2}}}}

\newcommand{\mV}[1]{\measValue{#1}{mV}}
\newcommand{\V}[1]{\measValue{#1}{V}}

\newcommand{\mHz}[1]{\measValue{#1}{mHz}}
\newcommand{\Hz}[1]{\measValue{#1}{Hz}}
\newcommand{\kHz}[1]{\measValue{#1}{kHz}}
\newcommand{\MHz}[1]{\measValue{#1}{MHz}}
\newcommand{\GHz}[1]{\measValue{#1}{MGz}}

\newcommand{\MeV}[1]{\measValue{#1}{MeV}}
\newcommand{\GeV}[1]{\measValue{#1}{GeV}}

\newcommand{\MeVcm}[1]{\measValue{#1}{MeV/cm}}

\newcommand{\MB}[1]{\measValue{#1}{MB}}
\newcommand{\GB}[1]{\measValue{#1}{GB}}

\newcommand{\Mj}[1]{Mj={#1}}

\newcommand{\about}{\ensuremath{\approx}}


\abstract{
  The ICARUS liquid argon TPC detector is taking data on the Booster (BNB) and Main Injector (NuMI) Neutrino beam lines at Fermilab with a trigger system based on the scintillation light produced by charged particles in coincidence with the proton beam extraction from the accelerators.
  The architecture and the deployment of the trigger system in the first two runs for physics are presented, as well as the triggered event rates. The event recognition efficiency has been evaluated as a function of the deposited energy and the position of cosmic muons stopping inside the detector.
}

\arxivnumber{2506.20137}

\maketitle
\flushbottom
    
\section{Introduction}

    The ICARUS-T600 liquid argon (LAr) time projection chamber (TPC) detector is currently installed and taking data at shallow depth with the Booster (BNB) and Main Injector (NuMI) Neutrino Beams since 2021.
    It plays the role of far detector in the Short Baseline Neutrino (SBN) program at Fermilab (US), searching for a possible LSND-like sterile neutrino signal at $\Delta m^{2} \about 1\,\textnormal{eV}^{2}$ \cite{acciarri2015proposaldetectorshortbaselineneutrino}.

    The detector consists of two identical modules, \m{19} long, \m{3.6} wide and \m{3.9} tall each, filled with 760\,t of ultra-pure liquid argon. A detailed detector description can be found in~\cite{Abratenko2023}. Each module houses two LAr TPCs separated by a 58\%-transparent shared cathode with a maximum drift distance of \m{1.5}, equivalent to \ms{\about\!1} drift time for the nominal 500\,V/cm electric drift field. The four TPCs (53,248~wires in total) are each characterized by three parallel readout wire planes with \mm{3} pitch, placed \mm{3} apart from each other and oriented at 0 and \textpm60\textdegree{} with respect to the horizontal direction. By appropriate voltage biasing, the first two planes (Induction~1 and Induction~2) provide a nondestructive charge measurement, whereas the ionization charge is fully collected by the last (Collection) plane.
    
    Scintillation light emitted by liquid argon when traversed by ionizing particles is detected by 360 8"~photomultiplier tubes (PMT) installed behind the TPC wire planes and used also for triggering purposes.

    Operating at shallow depth exposes the detector to an abundant flux of cosmic rays which could overwhelm detector event reconstruction during the 1 ms time window required to drift the ionization electron to the TPC wires. 
    To cope with this challenging condition, the detector is shielded by a \m{2.85} concrete overburden and surrounded by a segmented Cosmic Ray Tagger (CRT) system (about \sqm{1100}), which tags the surviving incoming charged particles. The CRT is composed of two layers of fiber-embedded plastic scintillators equipped with silicon photomultipliers, installed under the overburden (``Top CRT'') and externally on the four sides of the TPCs (``Side CRT'')~\cite{Aduszkiewicz_2025}.
    
\begin{figure}\centering
        \includegraphics[width=0.495\linewidth]{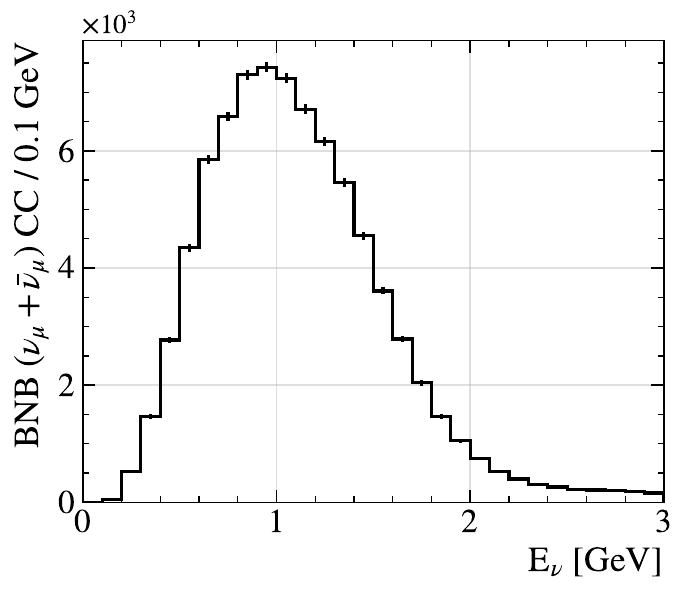}
        \hfill
        \includegraphics[width=0.495\linewidth]{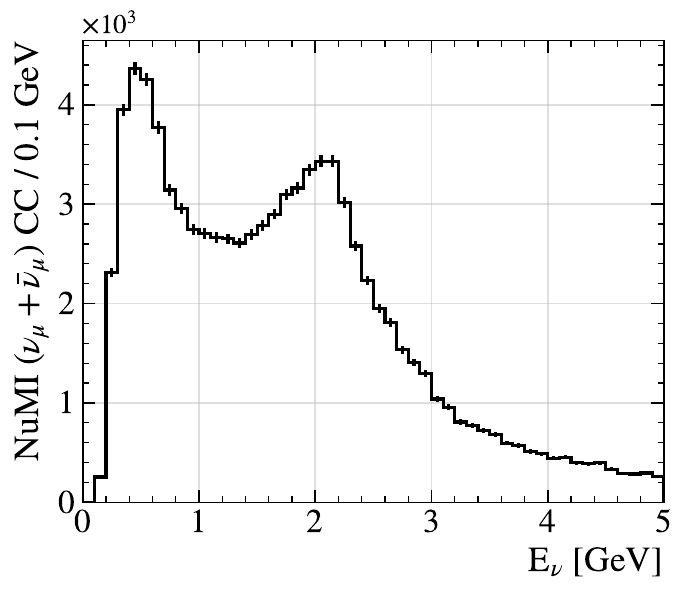}
        \caption{Charged current interaction spectra for muon neutrinos and anti-neutrinos in the active volume of the ICARUS detector, as determined in the BNB (\emph{left}) and NuMI off-axis (\emph{right}) beam calculations used in \cite{acciarri2015proposaldetectorshortbaselineneutrino, icaruscollaboration2024searchhiddensectorscalar} for $10^{20}$ POT.}
        \label{fig:NeutrinoSpectra}
    \end{figure}

    The main ICARUS trigger exploits the coincidence of the BNB and NuMI beams spills, \us{1.6} and \us{9.5} respectively, with the prompt scintillation light signals detected by the PMT system. Due to the \GeV{<2} (\GeV{<5}) neutrino energy range of BNB (NuMI) beams (see \cref{fig:NeutrinoSpectra}), the bulk of neutrino interaction events is expected to be contained in about 4~m along the beam direction, supporting the choice of a trigger based on the identification of fired PMTs inside a limited TPC region \cite{PMTS:Ali-Mohammadzadeh_2020}. 
    The data taking is complemented by routinely collecting events without any request on the scintillation light, with and without the presence of the beam, and also with high-rate external triggers for calibration purposes.

    The present paper is devoted to the description and characterization of the trigger system deployed for the first ICARUS data taking for physics with BNB and NuMI neutrino beam in the periods June~9 -- July~14, 2022 (``\RunOne'') and December~20, 2022 -- June~10, 2023 (``\RunTwo''). 
    The instrumentation adopted for the trigger signal generation starting from the PMT signals distribution, the accelerator signals of proton on target extraction and the required time synchronization system are described in \cref{sec:hardware}, as well as the light-based triggering logic and the specific triggers implemented. 
    A detailed study of the trigger performance in terms of event recognition efficiency is performed with cosmic rays, as described in \cref{sec:trig_eff}.

\section{The ICARUS trigger system}
\label{sec:hardware}

The ICARUS trigger system provides a global trigger signal which activates \ms{1.6} and \us{26} acquisition windows for TPC and PMT signal recording respectively. 
The duration of these windows is driven by the need to collect all the drift charge (the maximum drift duration for the \m{1.5} drift is $\about$1 ms) and all the scintillation light signals (more details in section \ref{ssec:hardware:PMT}) for each interaction event.
For each global trigger and within a \ms{2} time window (``enable gate'') around the beam spill, additional trigger signals (``local PMT trigger primitives'') can also be generated in coincidence with scintillation light flashes to activate the recording of \us{10} PMT waveforms, useful to identify the timing of cosmic muons crossing the ICARUS TPCs during the drift time.

\begin{figure}
    \centering
     \includegraphics[width=1\linewidth]{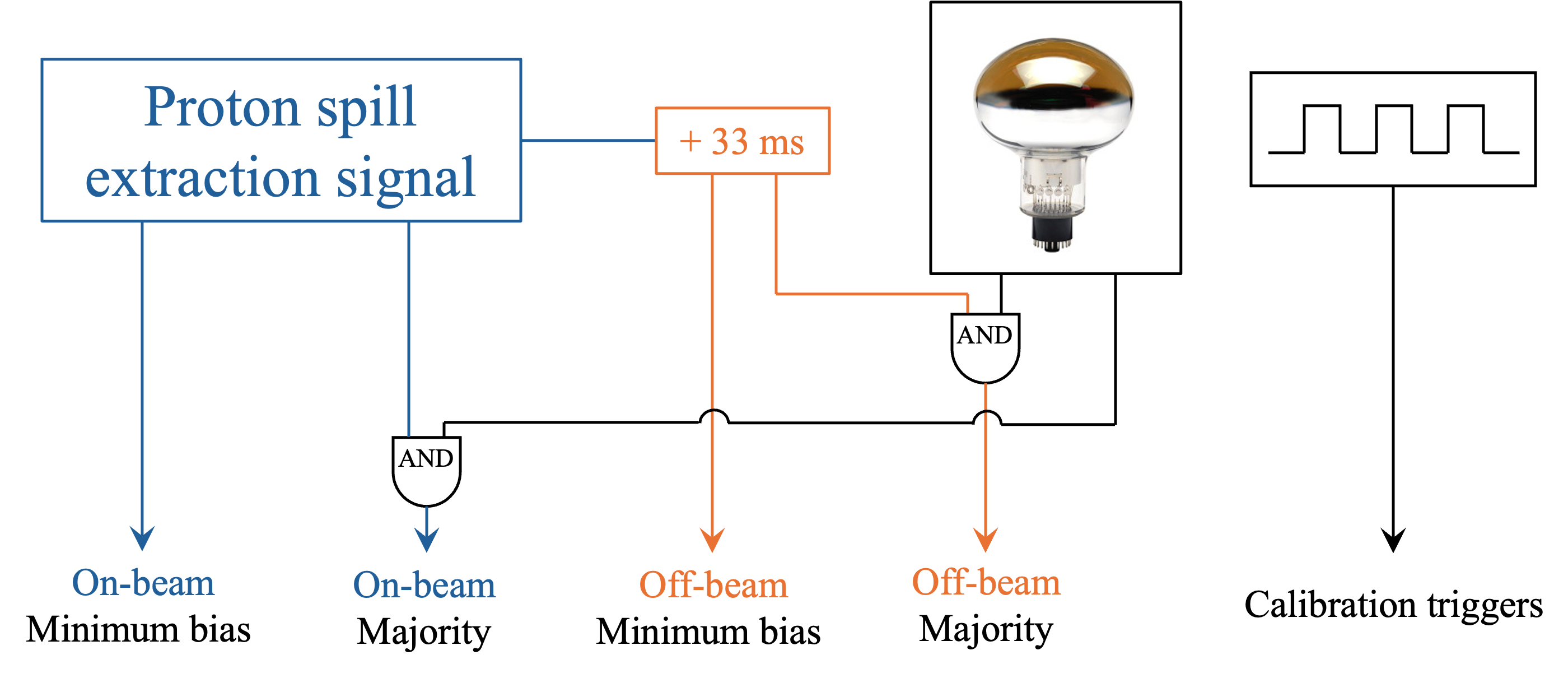}
     \caption{Schematics of the different mechanisms for trigger generation in the ICARUS detector.}
    \label{fig:Trigger_menu}
\end{figure}

The trigger architecture allows for the acquisition of different types of events, with and without the presence of the neutrino beams (see \cref{fig:Trigger_menu}). The main ICARUS global trigger (``on-beam majority'') is based on the multiplicity of PMT signals in coincidence within a ``beam gate'' opened at the expected arrival time of neutrinos as provided by the proton spill extraction signals of BNB and NuMI beams (see \cref{sec:beams}).
Off-beam cosmic ray events are collected with similar ``off-beam majority'' global triggers, based on the coincidence between light signals and off-beam gates generated 33~ms after each on-beam gate, away from the beam spills.
In addition, on-beam and off-beam ``minimum bias'' global triggers are generated in the presence of the corresponding gates, regardless of the scintillation light signals, to provide an unbiased data sample.
Minimum-bias triggers are collected once every 20 off-beam gates, and every 200 (60) on-beam gates for BNB (NuMI), to minimize the impact on triggered beam data taking,
but to still collect an adequate sample of unbiased triggers.
Unbiased data can be used to validate the performance of the majority trigger, for detector physics studies, and calibrations. 
The ICARUS trigger system can also operate in a multi-Hz ``calibration'' mode when the accelerators are not operational, opening gates synchronized to a pulse generator, to collect off-beam data at a much higher rate than during beam operations.

\subsection{Hardware setup and implementation}
\label{ssec:hardware:PMT}

\begin{figure}[!ht]
    \centering
     \includegraphics[width=0.33\linewidth,height=3in]{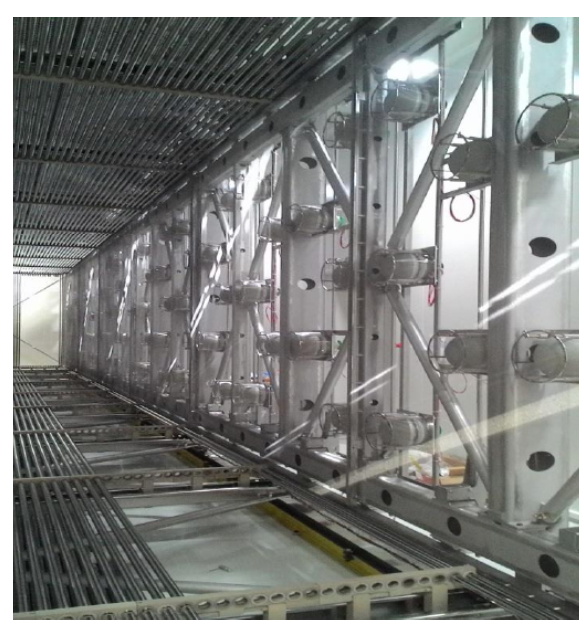} 
     \includegraphics[width=0.66\linewidth, height=3in]{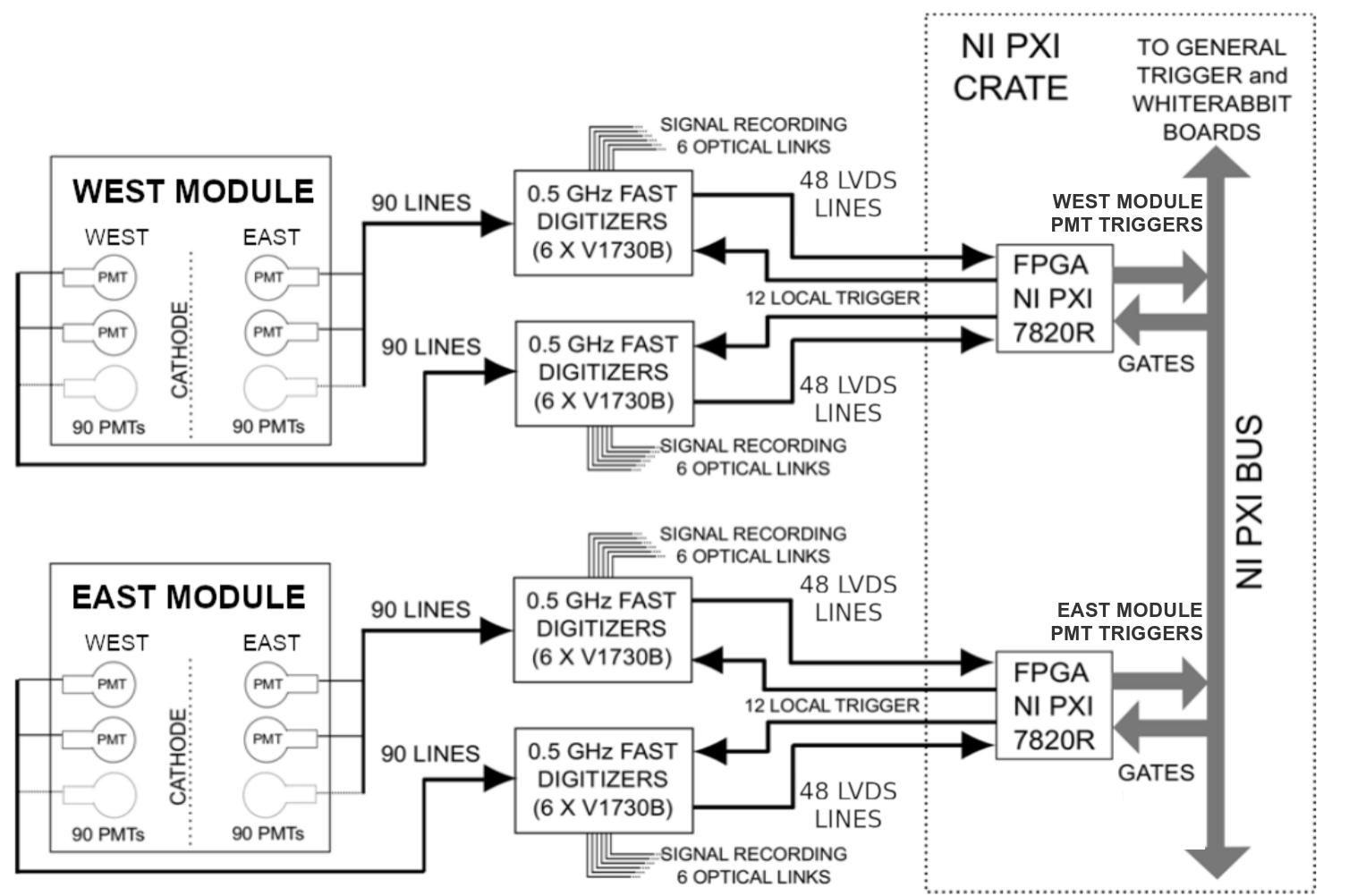} 
     \caption{(\emph{Left}) PMTs installed behind the wires in a TPC of one ICARUS cryostat. (\emph{Right}) Schematics of the PMTs readout electronics.}      
    \label{fig:PMT_electr}
\end{figure}

The scintillation light is collected by 360 Hamamatsu R5912-MOD 8''~PMTs made sensitive to vacuum ultraviolet (VUV) photons by 200\,\textmu{}g/cm$^2$ tetraphenyl butadiene coating deposited on their external window~\cite{PMTS:Ali-Mohammadzadeh_2020}.
The PMTs are installed behind the wire planes of each of the 4 TPCs, with 90 PMTs per TPC (\cref{fig:PMT_electr}). 

The signals coming from the 360 PMTs are read out by 24 CAEN V1730B digitizers, installed in 8 VME crates, 3 boards per crate.
Each digitizer hosts 16~flash~ADC channels, with 14-bit {500-MS/s} sampling and 2\,V input dynamic range.
In each board 15 channels are used for the acquisition of PMT signals, while the 16th channel is used to record three reference timing signals: the global trigger signal, the proton spill extraction signal, also used to generate beam gates (\cref{sec:beams}), and the signal of actual arrival of protons on the beam target from the resistive wall monitor counters~\cite{RWM}. This last signal enables physics analyses to determine the crossing time of beam neutrino bunches with nanosecond precision~\cite{Vicenzi:2025bvh}.

The digitizers generate trigger-request logical patterns via low voltage differential signal outputs (LVDS), which are activated by the presence of PMT signals with amplitude exceeding a digitally programmed threshold. Each digitizer produces 8 LVDS signals, 7 OR combinations of PMT pairs and one PMT singlet,
that are fed into the hardware modules that implement the trigger logic.

\begin{figure}
   \centering
   \includegraphics[width=1\linewidth]{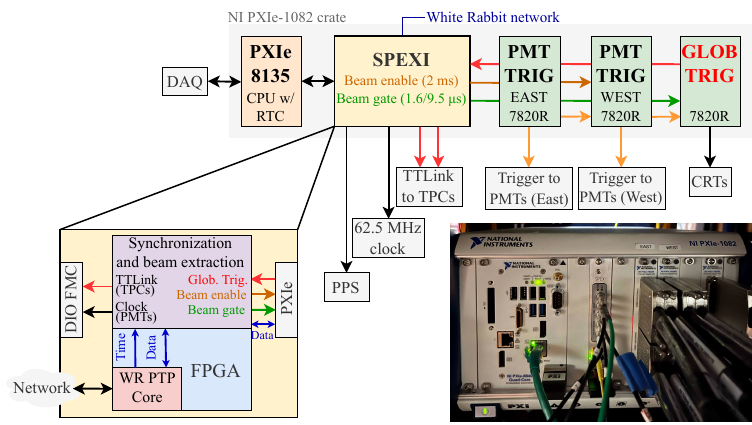} 
   \caption{(\emph{Left}) Schematics of the trigger system architecture layout and the main SPEXI functions as indicated in the expanded view. (\emph{Right}) Photo of the NI-1082 PXIe crate installed in a rack on the east mezzanine of the SBN-FD building. }
   \label{fig:trigger_crate}
\end{figure}

The trigger hardware setup (\cref{fig:trigger_crate}) consists of a single NI~PXIe-1082 crate containing a NI~PXIe-8840 real time controller (RTC), one mezzanine carrier board in PXIe form factor by INCAA Computers (SPEXI) and three NI~PXIe-7820R Field-Programmable Gate Arrays (FPGAs).
The SPEXI board generates signals based on the information of the proton extraction at BNB or NuMI through the White Rabbit (WR) network (\cref{sec:beams}). 
Such signals are: (i) a pulse-per-second (PPS) used both to reset the PMT digitizer time counters and as one of the reference signals of the CRT Front-End Boards \cite{Aduszkiewicz_2025}; (ii) \MHz{62.5} clock for the PMT digitizers; (iii) \MHz{10} clock and trigger for the TPC digitizers via a serial link named TTLink \cite{CAEN:a2795-manual}; (iv) \us{2.2}-wide (\us{10.1}-wide) beam gate for BNB (NuMI); (v) \ms{2}-wide enable gate for activating the PMT waveform readout outside of the beam spill.
While the clock signals are output on the front panel, the beam gate signals are distributed through the back-plane of the trigger crate to be available to the other modules for further use. 

Two of the three FPGAs, one per cryostat, are dedicated to processing the LVDS signals from the PMT digitizer boards to generate PMT trigger primitives according to the logic described in \cref{sec:trig_logic}. Each trigger primitive is distributed both on the back-plane of the trigger crate and to the PMT digitizer boards for PMT signal recording when in coincidence with the beam enable gate window. The third FPGA combines all signals present on the back-plane (beam gates and PMT trigger primitives) to produce the global trigger,  and vetoes the generation of other trigger primitives for the rest of the beam gate to ensure having only one global trigger for each spill.

The global trigger signal is distributed both externally to the CRT system and via the backplane to the SPEXI board that records its White Rabbit timestamp into a register and encodes it in the TTLink signal sent to the TPC crates. In addition, the third FPGA sends an Interrupt Request signal to the RTC, for reading all registers in the SPEXI and the FPGAs and assembling the global trigger information (timestamps, gate ID, LVDS states) into a TCP/IP packet sent to the data acquisition system (DAQ), which uses this timestamp to assemble all the subsystem data into an event.

The signals from PMTs are continuously digitized and recorded in circular memory buffers of \us{10} inside the V1730B boards. 
Memory buffers are set to include $3$ and \us{7} of samples before and after each PMT trigger primitive signal, respectively, to calculate a reference baseline of the PMT digitizer signal, and to collect both de-excitation components of the LAr scintillation light (fast $\tau \about \ns{6}$ and slow $\tau \about \us{1.6}$).
Furthermore,
in case another trigger primitive is issued within the \us{10} PMT acquisition window, the readout is extended to include \us{7} after the second trigger.
To record more scintillation light activity in correspondence of the global trigger, the PMT acquisition window for triggers in coincidence with the beam gate is enlarged to \us{26} (\cref{fig:timing}), starting from \us{7} before the beam gate opening.

When a PMT digitizer receives a trigger primitive, it generates another timestamp (trigger time tag, TTT) that is written in the header of the PMT data fragments. The TTT comes from an internal 31-bit counter, resets every second to avoid uncontrolled overflows using the PPS signal from the SPEXI, and is characterized by a \ns{16} time resolution. As a result, the timestamp associated to the PMT waveforms represents the number of nanoseconds elapsed since the last full second of GPS time.   
In order to improve the synchronization between trigger and PMT readout, a copy of the global trigger signal is sampled in the 16$^{\text{th}}$ channel of one digitizer in each crate, and its time is used as an external reference to correct the timestamps event by event.
The time of the PMT pulses is then reconstructed by combining together the timestamp of the global trigger from the White Rabbit system and the TTT, corrected by the 
relative time of the global trigger signal within the digitized window.

The CRT modules operate in a dedicated self-trigger mode~\cite{Aduszkiewicz_2025} and record the time with \ns{1} sampling, using as reference both the PPS generated by the SPEXI board, and the global trigger signal.
As a result, crossing particles are tagged by the CRT system with a time resolution at the level of 5 ns.

The precise synchronization between the CRT and the PMT systems is enabled by sharing the global trigger signal, used as timing reference for both reconstructed scintillation light signals and CRT signals as just described.

\subsection{Accessing the proton accelerator signals}
 \label{sec:beams}
 
The generation of the beam and enable gates used for triggering relies on the timely reception of the beam ``early warning'' signals provided by the accelerator complex to inform about the expected extraction of protons to BNB and NuMI targets~\cite{miniboone_booster_description}. Two such signals are made available at each of the two beam locations (MI-12 for BNB and MI-60 for NuMI).

The two early warning signals for BNB are the \texttt{\$1D}, which arrives \ms{35} before the extraction of the protons to the BNB target and is used to generate the enable gate, and the \texttt{gatedBES}, arriving \ms{0.3} before the beam extraction and used to generate the beam gate.
The \texttt{gatedBES} signal is generated as the coincidence between the \texttt{BES} (beam extraction signal) and the \texttt{\$1D} signals, to ensure the gate is opened only when the proton extraction is directed to the BNB beam line.
The two corresponding  signals for the NuMI beam are the \texttt{\$AE}, arriving \ms{730} before the beam extraction and used to open the enable gate, and the \texttt{MIBS\$74} arriving \ms{1.75} before the beam extraction and used to generate the beam gate.

\begin{figure}
    \centering
    \includegraphics[width=1\textwidth,trim=10 0 10 0,clip]{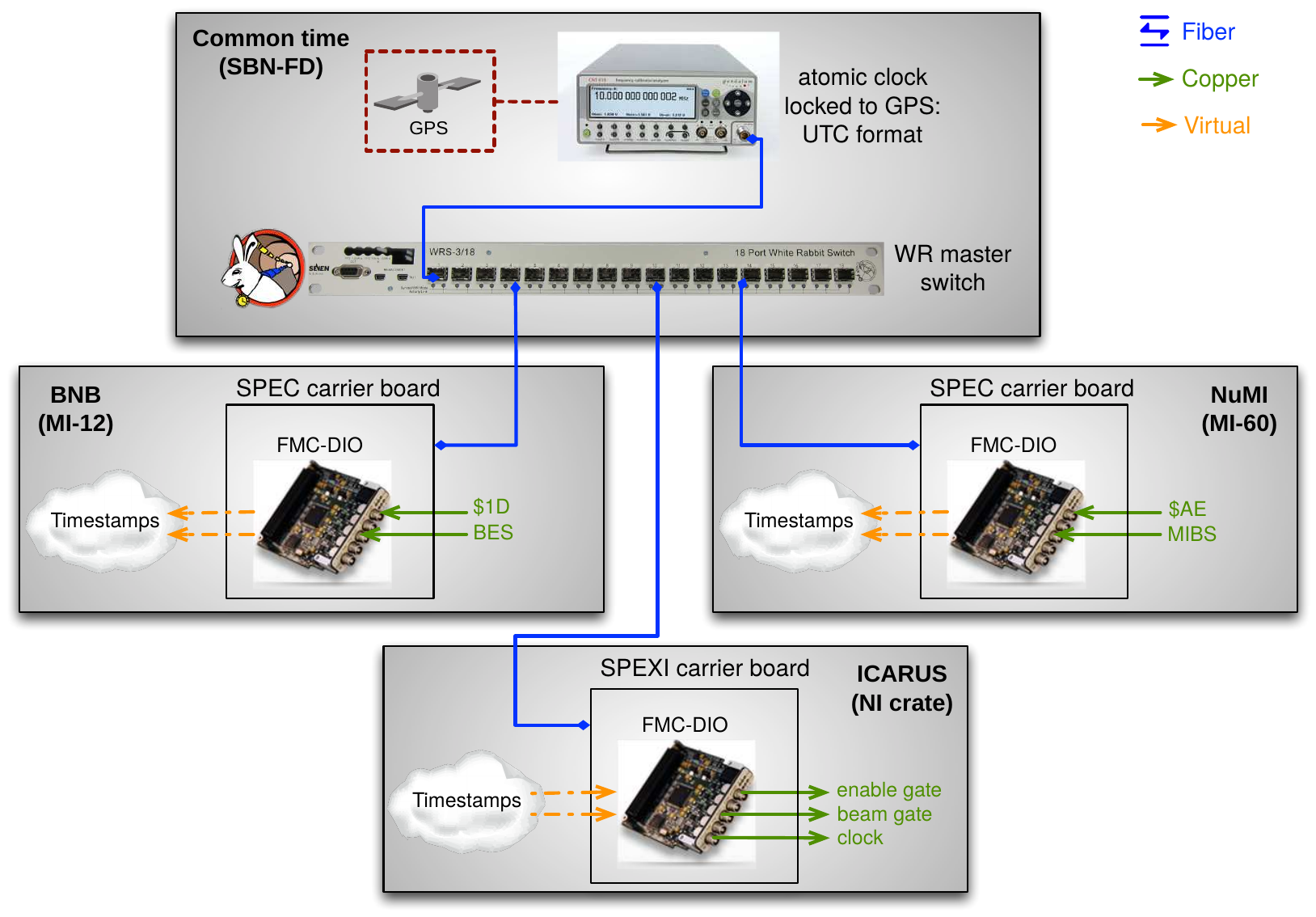}
    \caption{
    \label{fig:WR}
    Layout of the White Rabbit network implemented for the distribution of beam signals from the beam locations to the ICARUS building.
    }
\end{figure}

Distribution from the beam locations to the ICARUS building (SBN-FD) occurs via White Rabbit (WR), an Ethernet based network for synchronization of distributed systems with sub-nanosecond accuracy and precision better than \ps{50}~\cite{WhiteRabbit}. It consists of one master switch, disciplined by a GPS-locked atomic clock, connected to three nodes: 2~PCI-e cards (SPEC) in computers at locations MI-12 and MI-60, and the aforementioned SPEXI card in the NI crate at SBN-FD (see \cref{fig:WR}). 

\begin{figure}
    \includegraphics[width=0.49\textwidth]{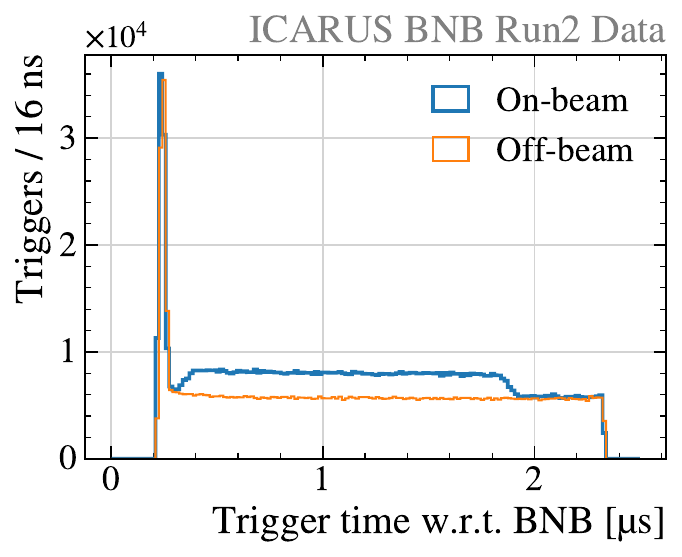}
    \hfill
    \includegraphics[width=0.49\textwidth]{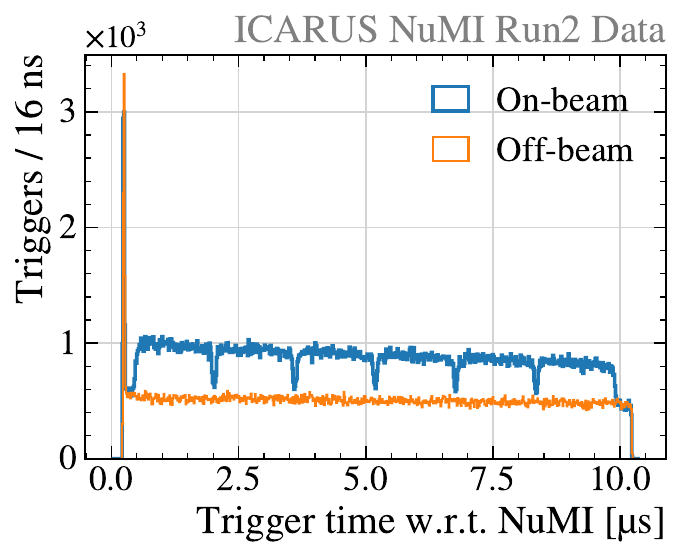}
    \caption{
    \label{fig:NeutrinoExcess}
    Profile of trigger time with respect to the beam gate opening time for the BNB (\emph{left}) and NuMI (\emph{right}) beams.
    The excess due to neutrino interactions is visible between \us{0.4} and \us{2.0} and between \us{0.4} and \us{9.9}, respectively.
    The sharp peak at the opening of the beam gate is caused by cosmic rays passing earlier than the gate opening, for which some late component of the scintillation light is emitted within the beam gate.
    }
\end{figure}

The SPEC WR nodes perform the timestamping of the leading edge of all the early warning signals and broadcast them over the WR network with \ns{8} resolution using Digital Input Output mezzanine (FMC-DIO) and Network Interface Card (NIC) core. The SPEXI receives such information and uses it, with appropriate time offsets, to open the enable and beam gates. These time offsets were validated and fine-tuned with beam data. The duration of the beam gates (2.2 and \us{10.1} for BNB and NuMI) is programmed to be larger than the duration of the beam extraction (1.6 and \us{9.5} respectively) to account for possible jitters. The excess of neutrinos and beam-associated particles over the cosmic rays, clearly identified inside the trigger beam gates for both BNB and NuMI in the collected data (\cref{fig:NeutrinoExcess}), confirms the synchronization of the trigger gates with the neutrino beam spills.

\subsection{Implementation of the PMT-majority logic}
\label{sec:trig_logic}

\begin{figure}
\centering
\includegraphics[width=\textwidth]{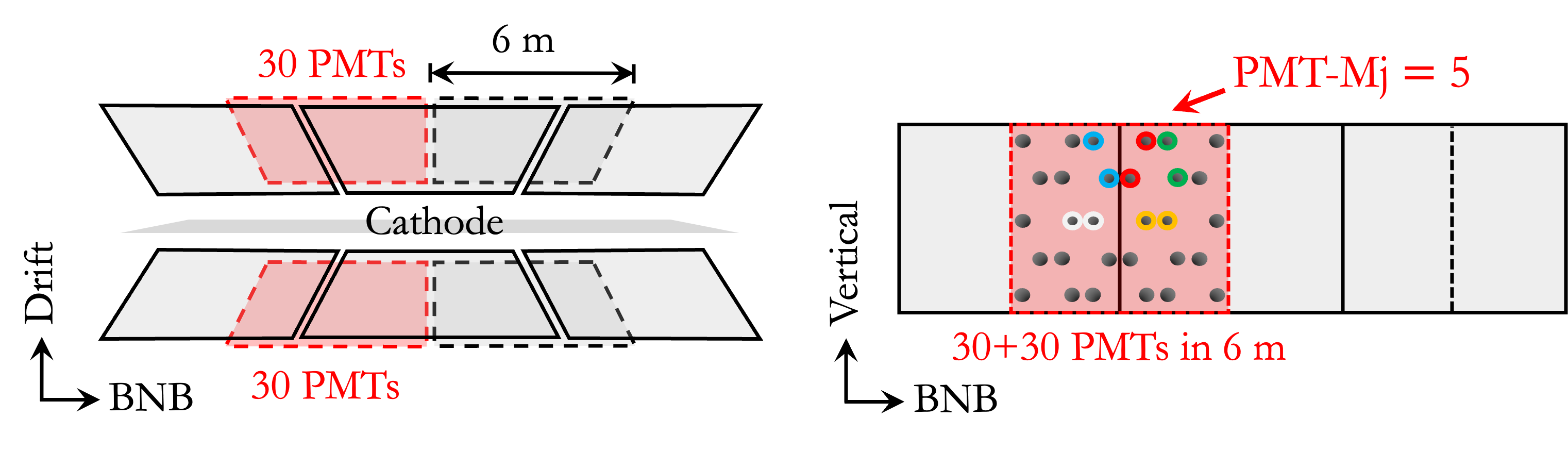}
\caption{(\emph{left}) View from the top of an ICARUS module, with two LAr-TPCs separated by a common cathode. For the trigger logic, light signals are evaluated in three side-by-side \m{6} longitudinal module slices. For \RunTwo, two additional slices (dashed lines) were introduced to improve efficiency and uniformity. (\emph{right}) Side-view of the module. In the highlighted window, containing 60 PMTs (30 per TPC), an example of a \Mj{5} pattern is shown, with PMTs paired into LVDS channels.}
\label{fig:windows}
\end{figure}

The longitudinal span of each TPC was divided into three logical windows, each one \m{6} long and containing the signals of 60 PMTs (30 PMTs per TPC, on opposite sides of the central cathode), in which the multiplicity of OR-paired PMT signals exceeding a 390 ADC threshold (approximately 13 photoelectrons) is evaluated.
This initial configuration, deployed in \RunOne, was upgraded in \RunTwo{} by implementing two additional overlapping windows, shifted by half a window (\cref{fig:windows}), to improve the trigger uniformity along the beam direction.

\begin{figure}
\centering
\includegraphics[width=\textwidth]{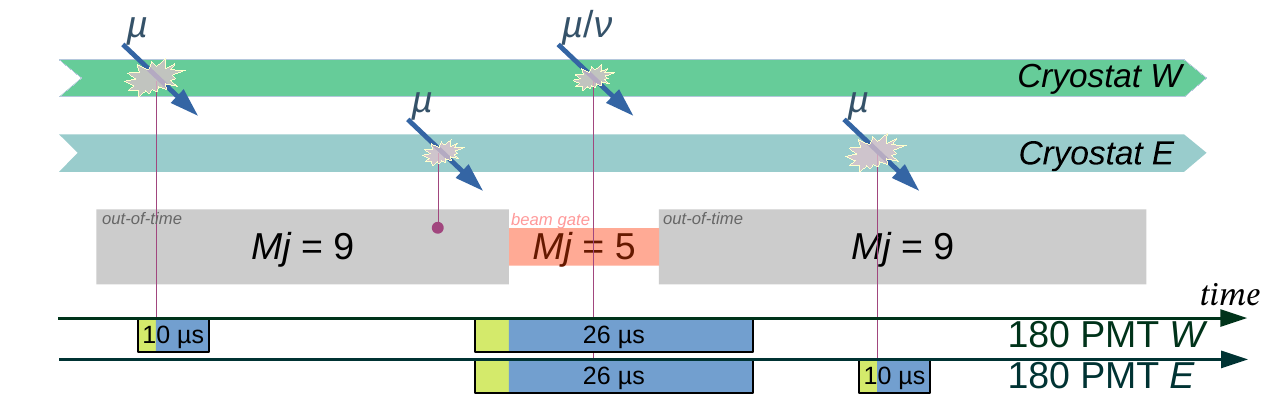}
\caption{Schematics of the logic implemented for issuing Global and local PMT trigger primitives and of the corresponding readout of the PMT waveforms. 
}
\label{fig:timing}
\end{figure}

For each \m{6} window, the multiplicity of LVDS channels from PMT pairs is evaluated on every \ns{25} FPGA clock cycle, independently by the two FPGAs dedicated to the east and west modules.
When at least 5 LVDS channels are fired within one window and inside the beam gate, a local PMT trigger primitive is generated in the corresponding module (\cref{fig:timing}) to evaluate the global trigger decision.
The choice of the majority levels was driven by the need of a high trigger efficiency for collecting BNB and NuMI neutrinos, while ensuring a sustainable data rate and size. 
In particular, looser conditions (at least 5 active LVDS channels, ``\Mj{5}'') were applied inside the beam gate to maximize the sensitivity to neutrino interactions with energy depositions as low as 200 MeV.
Tighter conditions, at least 10 and 9 active LVDS channels (\Mj{10} and \Mj{9}), were applied outside of the spills in Run1 and Run2 respectively, to limit the amount of collected PMT data while still being able to reject energetic cosmic rays in the analysis stage.

For the Run2 data taking, a typical trigger rate of \Hz{\about0.7} was obtained, including \Hz{\about0.2} and \Hz{\about0.25} respectively from the BNB and NuMI majority components and \Hz{\about0.25} from minimum-bias triggers (\cref{table:TriggerRatesRun2}). This resulted in a manageable data readout bandwidth with good operational stability.

\begin{table}
    \caption{
        \label{table:TriggerRatesRun2}
        Typical trigger rates observed during the \RunTwo{} data taking with the BNB and NuMI beams.
    }
    \begin{center}
        \begin{tabular}{cccccc}
            \hline
            beam &  \multicolumn{2}{c}{on-beam} & \multicolumn{2}{c}{off-beam} & total   \\
                 &  majority    &  minimum bias &  majority    &  minimum bias &         \\
            \hline
            BNB  &  0.10~Hz     &  0.02~Hz      &  0.07~Hz     &  0.20~Hz      & 0.39~Hz \\
            NuMI &  0.15~Hz     &  0.01~Hz      &  0.08~Hz     &  0.04~Hz      & 0.28~Hz \\
            \hline
        \end{tabular}
    \end{center}
\end{table} 

\subsection{Collection of the triggered events}
\label{sec:collection_data}

The ICARUS data acquisition system and the trigger system communicate via TCP/IP protocol, transferring the trigger information generated by the SPEXI/FPGA boards as a data packet to the DAQ. The general ICARUS data acquisition architecture, based on the \emph{artdaq} framework developed at Fermilab~\cite{biery2018flexible}, is described in~\cite{Abratenko2023}.

Among other information, the trigger packet contains the gate opening and trigger timestamps, the gate type (on-beam or off-beam), the trigger type (majority or minimum bias trigger),
and the state of each of the LVDS channels at trigger time, providing PMT-level information on where the trigger occurred within the detector.
The inclusion of the \ns{8}-precise timestamps from the White Rabbit system allows to properly align the beam gate with the beam spills (see \cref{fig:NeutrinoExcess}).

The DAQ decodes the data packet sent by the real-time controller and places the information into a trigger ``data fragment'' stored in the corresponding event. The event building is driven by the presence and timestamp of a global trigger data fragment during data taking, to assemble all detector data into an event.

The beam intensity, i.e. the number of protons on target (POT) in each beam spill, is stored in dedicated databases hosted at Fermilab for both the BNB and NuMI beams. 
The spill-level POT information is correlated with the stored trigger information, allowing to evaluate the exposure associated to each event.

\section{Study of the ICARUS trigger efficiency}
\label{sec:trig_eff}

The measurement of the ICARUS PMT trigger efficiency takes advantage of the large cosmic ray dataset from multi-Hz rate minimum bias runs, collected in absence of the neutrino beam, and with no requirement on the scintillation light signals to remove any bias from the triggering logic itself.
The abundance of cosmic rays traversing all regions of the detector allows to investigate the detector response to the liquid argon scintillation light and to measure the trigger efficiency at different event positions inside the detector.
The PMT readout window was enlarged from \us{26} to a \us{166} interval to increase the rate for collecting large unbiased cosmic ray statistics.

The response of the trigger to the passage of each single cosmic ray was determined from the PMT waveforms recorded at the time the particle was crossing the detector, via dedicated software that emulates the implemented triggering logic.

\subsection{Timing of cosmic ray tracks in the TPC}
\label{ssec:TrackTimes}

The study of the trigger efficiency requires the knowledge of the crossing time ($t_0$) of each particle to unambiguously identify the tracks associated to the event.
Tracks from cosmic rays are reconstructed from TPC wire signals with the Pandora toolkit~\cite{Marshall2015},
commonly used in large LAr-TPC-based experiments~\cite{Acciarri2018,Abud2023}.
In general, the position of a ionizing particle along the drift direction is reconstructed from the timing of collected charge in the TPC and the particle crossing time, which are bound together through the electron drift velocity ($v_{D} \about 1.6\,\text{mm/\textmu{}s}$).
The time can be assigned to reconstructed tracks either when
they cross the cathode of the TPCs, or in association to signals in the CRT system surrounding the two modules.

\paragraph{Cathode-crossing tracks.}
\label{paragrap:CathodeCrossingReco}

In the ICARUS detector, cathode-crossing tracks can be efficiently identified by matching the two separately reconstructed segments in the facing TPCs.
This matching pins down the actual position of the track, allowing to reconstruct its crossing time from the spatial shift along the drift direction that is needed to connect the two segments.
The resulting time resolution, driven by the quality of the track reconstruction near the cathode, is of the order of microseconds, dominated by 1\%-level non-uniformities of the drift electric field \cite{ICARUS:2024calibration}.

\paragraph{Matching with the Cosmic Ray Tagger system.}
\label{sssec:CRTmatch}

\begin{figure}\centering	
    \includegraphics[width=0.495\linewidth]{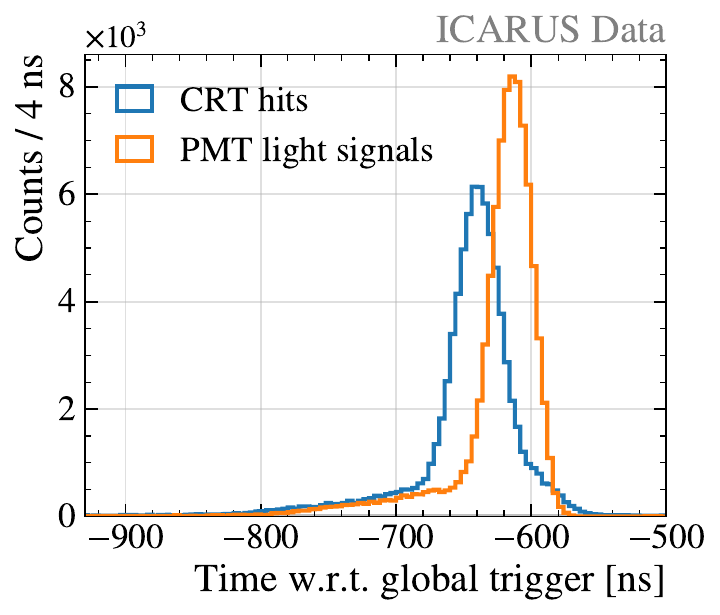}
    \hfill \includegraphics[width=0.495\linewidth]{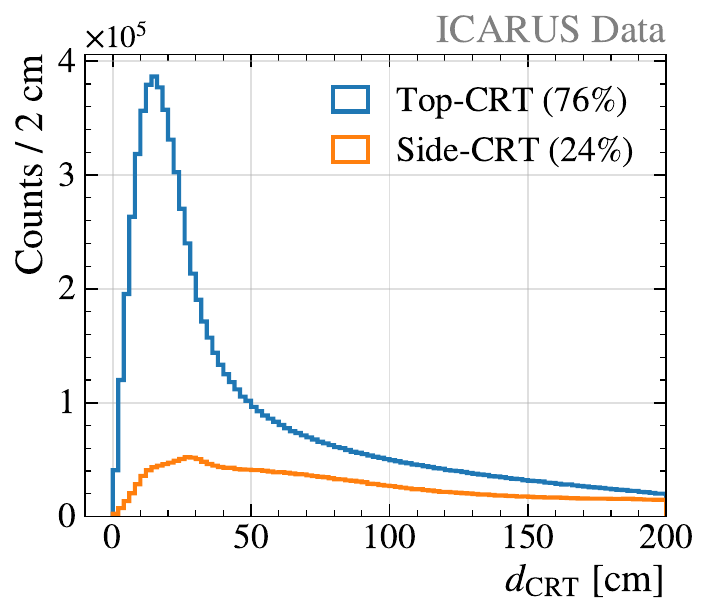}

\caption{(\emph{Left}) Time distribution of the CRT hits and scintillation light PMT signals close to the primary event trigger compared to the digitized global trigger signal.
The relative delay of the PMT signal reflects the additional time cosmic rays take from the CRT modules to the liquid argon,
in addition to the time of propagation of argon scintillation light to the PMTs.
The corresponding dataset consists of \about 90,000 off-beam BNB events collected during \RunOne{}.
(\emph{Right}) 
Distribution of the distance between the best-matched CRT hit to the reconstructed track position extrapolated to the corresponding CRT plane.} 
\label{fig:CRT-PMT-TPCmatch}
\end{figure}

The spatial and time correspondence of CRT signals with PMT light and TPC charge signals are used to reconstruct the time and position of the cosmic rays based on the corresponding signal on the CRT (``hit'').
For this purpose, the trajectory of each track is linearly extrapolated using its direction at both track ends to intersect the CRT counters.
Among the CRT hits whose time is compatible with the track position in the TPC, the hit with the smallest distance on the CRT plane (\dCRT{}) to any one of the two track-CRT intersections is chosen and its time is assigned to the track.
When applying the matching to the collected data, 76\% of the cosmic ray tracks are matched to the Top-CRT, and 24\% to the Side~CRT\footnote{The matching with $\dCRT < \m{1}$ has an estimated efficiency of $\about90\%$ for the Top~CRT and $\about50\%$ for the Side~CRT \cite{boone:2024fdy}.}(\cref{fig:CRT-PMT-TPCmatch}, right).

The PMT signal interval used to evaluate the response of the trigger to the track is determined by the time of the matched CRT hit.
The distribution of recorded times for CRT hits and PMT signals with respect to the digitized global trigger signal is shown in \cref{fig:CRT-PMT-TPCmatch} (left).

\subsection{Trigger emulation}
\label{ssec:TriggerEmulation}

The trigger response was studied with a software emulation of the trigger hardware operations based on the recorded PMT waveforms, to evaluate different trigger conditions with a single sample of unbiased data.
The emulation algorithm was applied in a \ns{300} window around the time of reconstructed tracks (\cref{ssec:TrackTimes}) to include the associated prompt scintillation light signals.
The algorithm replicated trigger hardware operations: discrimination of PMT signals at the leading edge and their pairing in OR logic, clustering in 3 or 5 logic windows (\cref{fig:windows}), counting of PMT-pairs to meet the required majority setting, and inferring the time the trigger would fire at. 

\begin{figure}\centering

    \includegraphics[width=0.65\linewidth]{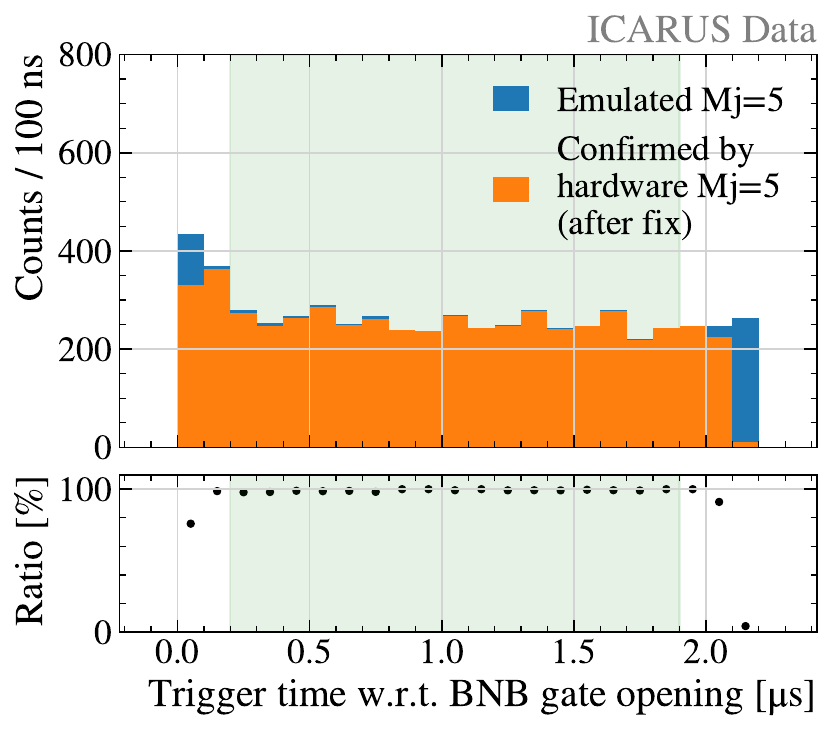}
    \caption{
        Emulated triggers in $<20 \ \text{h}$ minimum bias runs, after the correction to the PMT readout is introduced, so that PMT signals are discriminated at the leading edge in both the emulation and the hardware. 
        The fiducialized \us{[0.2, 1.9]} beam gate window is highlighted.}
    \label{fig:EmulatedTriggerTimeProfile}
\end{figure}

The consistency between emulation and hardware operation was validated on an event-by-event basis 
with a dedicated sample of minimum bias data that also included the information of the PMT-majority trigger at the hardware level.
The emulation was found to miss a trigger in less than 0.1\% of the hardware-triggered events, due to the unavoidable presence of small signal jitters in the hardware.
Conversely, the hardware trigger confirmed the emulated triggers at the level of $0.6\%$ (\cref{fig:EmulatedTriggerTimeProfile}) after a $\about5\%$ correction was introduced to fix an inconsistency between the misconfigured hardware and the software emulation, which discriminated PMT signals on the trailing and leading edge respectively (see \cref{ssec:EfficiencyMeasurement}).

\subsection{Analysis of the collected data} 
\label{subsec:data-analysis}

\begin{figure}\centering	
    \includegraphics[width=1\linewidth]{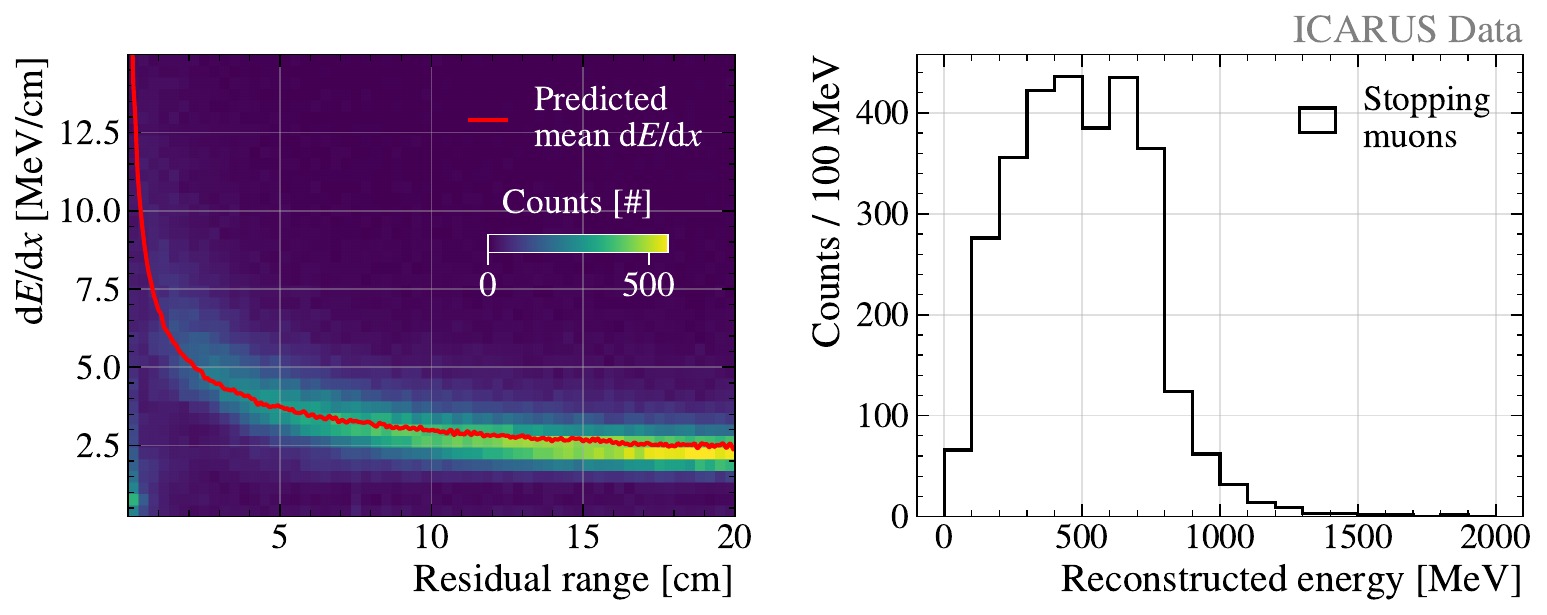}
    \caption{(\emph{left}) Distribution of \dEdx ionization with respect to the residual range for the selected stopping muons. The Bethe-Bloch mean prediction is shown for reference. (\emph{right}) Distribution of the energy reconstructed from the residual range. See the text for an in-depth description of the selection procedure.}
	\label{fig:stopping-muon-selection}
\end{figure}

A high-purity sample consisting of well reconstructed cosmic muons was selected to evaluate the trigger response as a function of the energy at different locations inside the TPC volume. 
Two main classes of cosmic ray events were considered:

\begin{itemize}
    \item Stopping muons as recognized by the increased \dEdx local energy depositions at the end of the tracks, to characterize the trigger as function of the energy. 
    They are uniformly distributed across the detector and are characterized by an energy spectrum in the same range as BNB charged-current muon neutrino interactions (see \cref{fig:NeutrinoSpectra,fig:stopping-muon-selection}), with roughly the same event spatial extension in the detector. 
    \item Almost vertical muons characterized by localized energy depositions in the longitudinal-drift plane of the detector, to study the uniformity of the trigger response with high granularity along the longitudinal direction and at different distances from the PMTs.
\end{itemize}
The event selection procedures were developed directly on data, maximizing the purity of the selected sample at the cost of a reduced statistics. Events were selected and reconstructed by matching tracks in the TPCs and associated CRT signals.
To minimize the impact of possible incorrect associations, the times assigned independently by CRT and TPC to cathode-crossing tracks must be consistent within \us{10}. 
Furthermore, to avoid spatial mismatch between the track and associated PMT activity, the presence of a weak PMT light signal (at least 40~photoelectrons integrated over \us{1} across at least 2~PMTs) was required within \m{1} of the track barycenter along the longitudinal direction. This very loose light requirement did not introduce significant biases in the measurement of the trigger efficiency even at low energy.
In addition, no other light signal was allowed in the previous \us{10}, to avoid including spurious signals.
\begin{figure}[ht!]\centering	
    \includegraphics[width=0.8\linewidth]{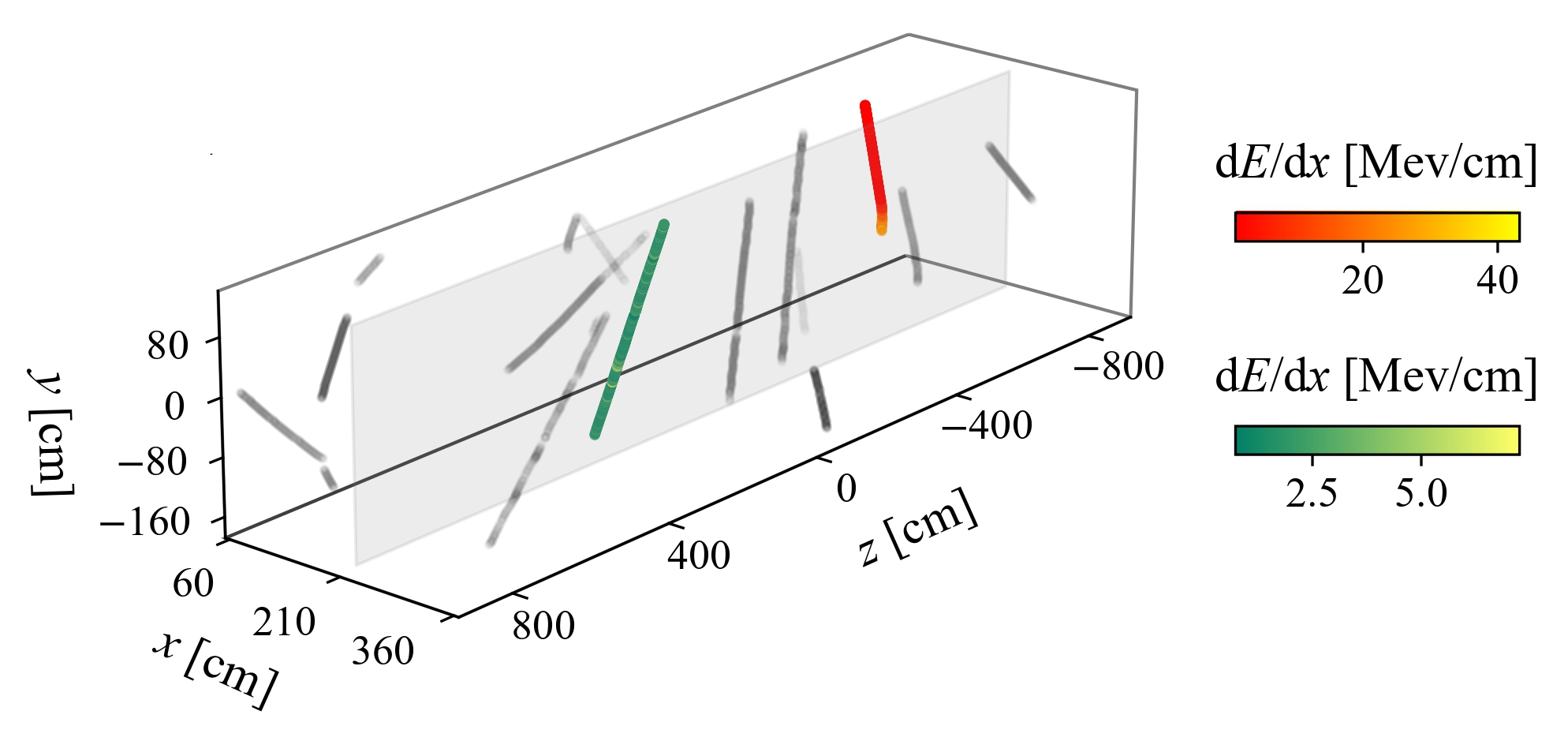}
    \caption{Example of event in the ICARUS West module from \RunTwo{} minimum bias data. The TPC is read out for \ms{1.64}, and the reconstructed tracks are matched with the CRT, which is read out for \ms{3}. Some of the tracks are clipped due to the TPC readout opening (closing) after (before) the particle enters the active volume of the detector. Tracks selected with the vertical and stopping selections are highlighted with green and red shades respectively.}
	\label{fig:event-display}
\end{figure}

Examples of selected through-going and stopping muons are visualized in \cref{fig:event-display}, with the corresponding \dEdx local energy depositions measured with granularity at millimeter level.
While the deposited energy for through-going muons is measured from the charge collected on the TPC wires with a $\about5\%$ resolution \cite{ICARUS:2024recombination}, the energy for stopping muons is more precisely determined from their reconstructed residual range.
For this purpose, the ionization charge was calibrated by accounting for the free drift electron lifetime in LAr (\ms{\about5} in the analyzed data sample) \cite{ICARUS:2024calibration}, and electron-ion recombination \cite{ICARUS:2024recombination}.

\paragraph{Stopping Muons.}
\label{par:StoppingMuons}

Muon tracks exceeding a length of \cm{20} were matched to CRT signals with $\dCRT < \cm{30}$, within the predefined \us{166} PMT readout window.
Tracks were required to originate within \cm{10} from the top or side TPC walls and to stop inside a ``fiducial volume'' defined by a \cm{10} padding from the TPC walls, and \cm{5} from the cathode.
Furthermore, a \cm{5} padding was introduced around the longitudinal center of the TPC where the split of Induction~1 wires may degrade track reconstruction.
The Bragg peak of the stopping muons was identified by requiring the median value of \dEdx ionization to be at least \MeVcm{4} in the last \cm{5}.
In addition, based on the Bethe-Bloch prediction for stopping muons in liquid argon, an energy deposition of at least \MeV{60} was required in the last \cm{20}.
 
\Cref{fig:stopping-muon-selection} (left) shows the distribution of the \dEdx with respect to the residual range for the selected stopping muons, along with the Bethe-Bloch prediction for the mean \dEdx value. The typical ascent at the end of each track is evident, highlighting the purity of the selected sample. 
The corresponding energy spectrum, measured from the reconstructed range, is reported in \cref{fig:stopping-muon-selection} (right).

\begin{figure}[b]\centering
    \includegraphics[width=0.45\linewidth]{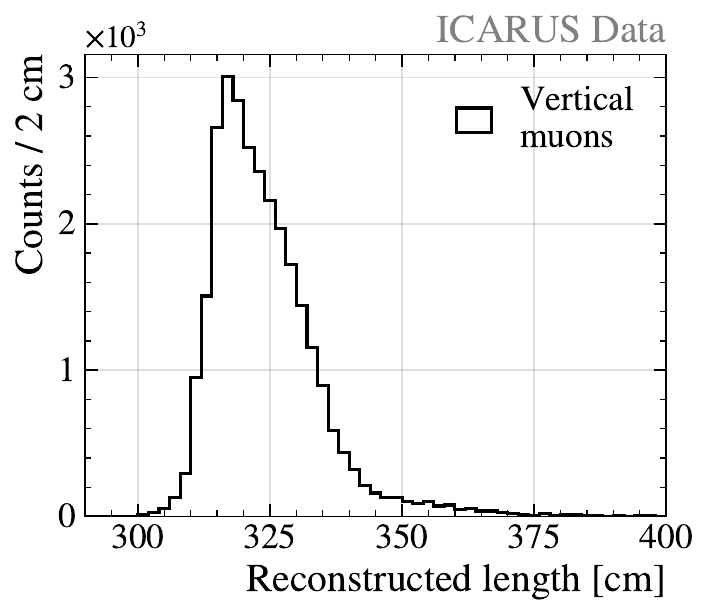}
    \caption{Reconstructed length for the selected through-going almost-vertical cosmic muons. See the text for an in-depth description of the selection procedure.}
	\label{fig:vertical-muon-selection}
\end{figure}

\paragraph{Vertical muons.}
\label{par:VerticalMuons}

Muon tracks longer than \cm{20} were matched to CRT signals with $\dCRT < \cm{30}$, within the PMT readout window.
A one-to-one match was required between each CRT hit and track, to reduce the contamination by tracks and showers reconstructed as multiple collinear tracks that are matched to the same CRT hit, which degrades the reconstruction of the tracks length and deposited energy.
In this way, muons were selected independently of their orientation (``loose selection'').\label{par:LooseSelection}
In addition, vertical muon tracks were selected entering the detector at a $< 20^{\circ}$ angle with respect to the vertical direction, and matched only to Top CRT signals.
These tracks were allowed to have the start (end) point up to within \cm{15} from the top (bottom) TPC wall to accommodate for distortions from electric field non-uniformity, and were required to have a reconstructed length in agreement within 3\% with the start-to-end distance.
The reconstructed length for the selected vertical muons is shown in \cref{fig:vertical-muon-selection}. 

\subsection{Trigger efficiency measurements}
\label{ssec:EfficiencyMeasurement}

The event recognition efficiency of the hardware-imposed PMT-majority logic was determined by the fraction of muon tracks meeting the software-emulated trigger condition out of the number of selected tracks, as a function of the measured energy and spatial position along the detector. 
The PMT signals were discriminated with a 400~ADC count threshold in \RunOne{}, and an equivalent threshold of 390~ADC counts in \RunTwo{} to take into account a slightly different detector configuration.

\begin{figure}\centering
    \includegraphics[width=0.495\linewidth]{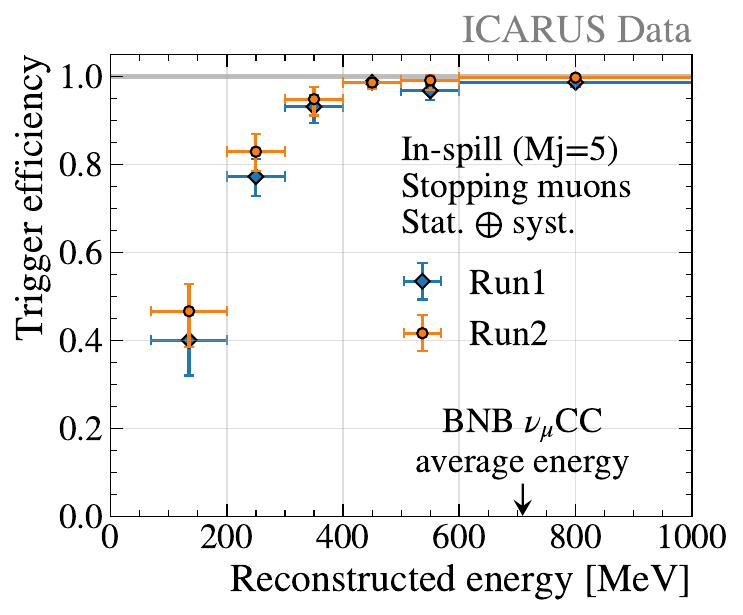}
    \hfill
    \includegraphics[width=0.495\linewidth]{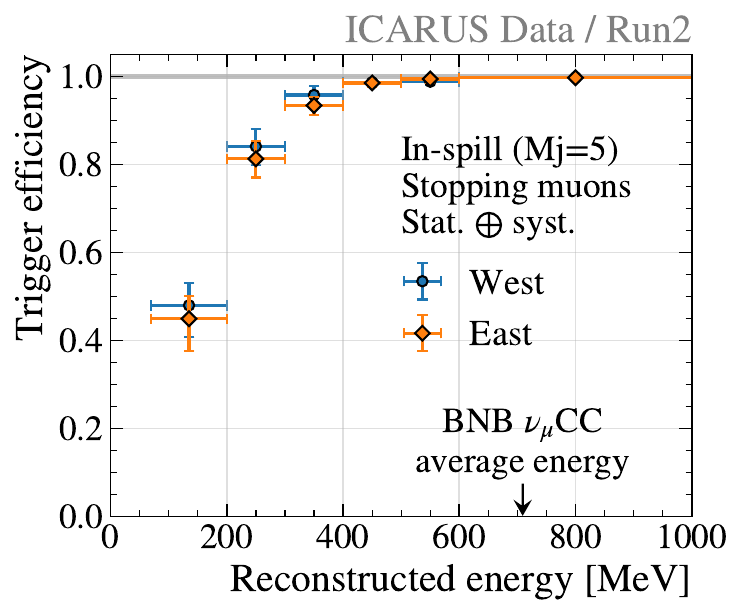}
    \caption{\Mj{5} trigger efficiency with stopping muons as function of the energy from the residual muon range in \RunOne{} and \RunTwo{} (\emph{left}), and separately in the two West and East modules for \RunTwo{} (\emph{right}). 
    The error bars include both systematic and statistic uncertainties added in quadrature. The average simulated deposited energy for BNB $\nu_{\mu}$ charged-current interactions is indicated for reference.}
	\label{fig:efficiency-energy}
\end{figure}

The majority-5 (\Mj{5}, see \cref{sec:trig_logic}) in-spill trigger efficiency on stopping muons is reported in \cref{fig:efficiency-energy} as a function of the muon energy determined from the residual range (\MeV{>70}).
The efficiency roughly saturates at \MeV{400} both in \RunOne{} and \RunTwo{}, with the plateau largely covering the spectrum of charged-current neutrino interactions from the BNB and the higher-energy NuMI beams (see \cref{fig:NeutrinoSpectra}). 
The introduction of the two overlapped windows in \RunTwo{} results in a $\about10\%$ improvement below \MeV{300} compared to \RunOne{}, where the detector was logically divided in only three regions (see \cref{fig:windows}).
The trigger efficiency in the East module shows a reduction by up to 3\% compared to the West module (\cref{fig:efficiency-energy}), due to the presence of 11 and 3 dead PMT channels in the two modules respectively.

Uncertainties on the measured trigger efficiency were evaluated with the same approach in both \RunOne{} and \RunTwo{}, where they are of similar magnitude. The statistical term ranges up to $\about6\%$ below \MeV{200} energy of the stopping muons, $\about2\%$ between $\MeV{200}$ and \MeV{300}, and $<1\%$ at higher energy. Associated systematics come from the track reconstruction and selection, and from the emulation of the trigger PMT-majority logic using the recorded PMT signals:
\begin{description}

  \item[CRT-TPC matching] (\cref{sssec:CRTmatch}). The trigger emulation software looks for PMT signals in a \ns{300} window around the time of the muon as assigned with the CRT. 
  A CRT-TPC mismatch would lead to a wrong evaluation of the track time. 
  This source of systematics was removed by requiring a loose spatial correspondence of a minimal light signal on few PMTs with the track charge deposition in the TPC. 
  In the case of cathode crossing tracks, the time assigned from the CRT was requested to match the one assigned from the TPC (see \cref{subsec:data-analysis}).
  
  \item[Reconstruction of muon tracks with the TPC.]

  The stopping muon selection, based on the identification of the Bragg peak by the measured ionization signal (see \mbox{\cref{subsec:data-analysis}}), allowed to correctly reconstruct the large majority of tracks, with a resolution on the length better than $1\%$ as indicated by a dedicated Monte Carlo study.
  In the few remaining cases where the reconstruction of the tracks is incomplete, tracks may be broken and wrongly recognized as stopping muons, e.g., due to the presence of kinks and delta rays. In this case, the track length and associated energy would be underestimated, leading to an overestimation of the trigger efficiency at low energies.
  The impact of track mis-reconstruction was evaluated via Monte Carlo and included in the measurement of the trigger efficiency as a lower-bound systematic uncertainty of 11\% below \MeV{200}, 1\% between 200 and \MeV{300}, and negligible everywhere else.
  \item[Software emulation of trigger hardware.]
  The software emulation of the trigger response, 
  applied to a dedicated minimum bias data sample in which additional trigger hardware information was added,
  showed $\about5\%$ of events with an emulated trigger not confirmed by the hardware.
  The tracks generating such emulated triggers were typically located at the corners of the detector and had low deposited energy, yielding weak and short light signals which might have been missed by the misconfigured hardware (see \cref{ssec:TriggerEmulation}).
  In \RunOne{}, the efficiency shows a reduction of $11\%$ below \MeV{200}, $5.5\%$ between 200 and \MeV{300}, $4.7\%$ between 300 and \MeV{400} and $<1\%$ at higher energies.
  In \RunTwo{}, the decrease is of $14\%$ below \MeV{200}, $8\%$ between 200 and \MeV{300}, $3.7\%$ between 300 and \MeV{400} and $<1\%$ above \MeV{400}.
  The trigger efficiency has been corrected accordingly, with a conservative systematic uncertainty amounting to half of the correction.
\end{description}

Overall, as shown in \cref{table:uncertainty-budget} the combined systematic uncertainties amount to $\about +7/-13\%$ below \MeV{200}, $\about4\%$ between 200 and \MeV{300}, $\about2\%$ between 300 and \MeV{400}, and are almost negligible at higher energies, where tracks are better recognized and measured.

\begin{table}
    \begin{center}
        \begin{tabular}{lcccc}
            \hline
                         & \MeV{<200} & \MeV{[200, 300]} & \MeV{[300, 400]} & \MeV{>400} \\
            \hline
            Statistical  & $6\%$      & $2\%$            & $<1\%$           & $<1\%$     \\
            \hline
            Systematic   & $+7\%/-13\%$ & $\pm4\%$       & $\pm 2\%$        &  $< \pm 1\%$ \\
            
                        \hline
        \end{tabular}
    \end{center}
    \caption{
        \label{table:uncertainty-budget}
        Breakdown of uncertainties on the \RunTwo{} trigger efficiency measurement as a function of the reconstructed energy, extracted from the sample of stopping muons. Similar values apply to \RunOne{}.
    }
\end{table} 

\begin{figure}\centering
    \includegraphics[width=\textwidth]{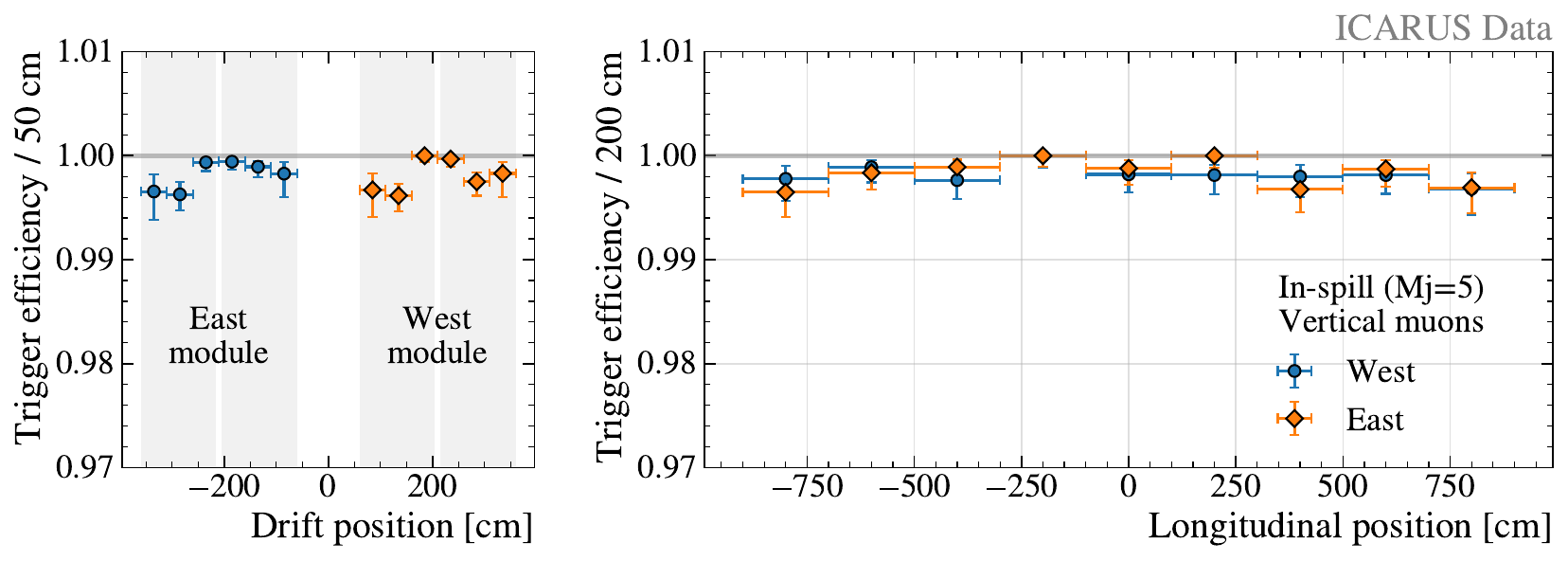}
    \caption{
    \RunTwo{} trigger efficiency with \Mj{5} (used to trigger on events inside the beam spills), measured with vertical muons in the \GeV{[0.6, 1.0]} energy range as function of the drift and longitudinal average position of the tracks.
    Only statistical errors are shown (systematics are negligible in this energy range).}
	\label{fig:efficiency-space}
\end{figure}

The spatial uniformity of the trigger response was probed with the selected vertical muons (see \cref{subsec:data-analysis}) in the \GeV{[0.6, 1.0]} deposited energy range.
The event recognition efficiency with the emulated \Mj{5} trigger is roughly saturated (\cref{fig:efficiency-space}) in both \RunOne{} and \RunTwo{}, regardless of the muon position. 

\begin{figure}\centering
    \includegraphics[width=1\textwidth]{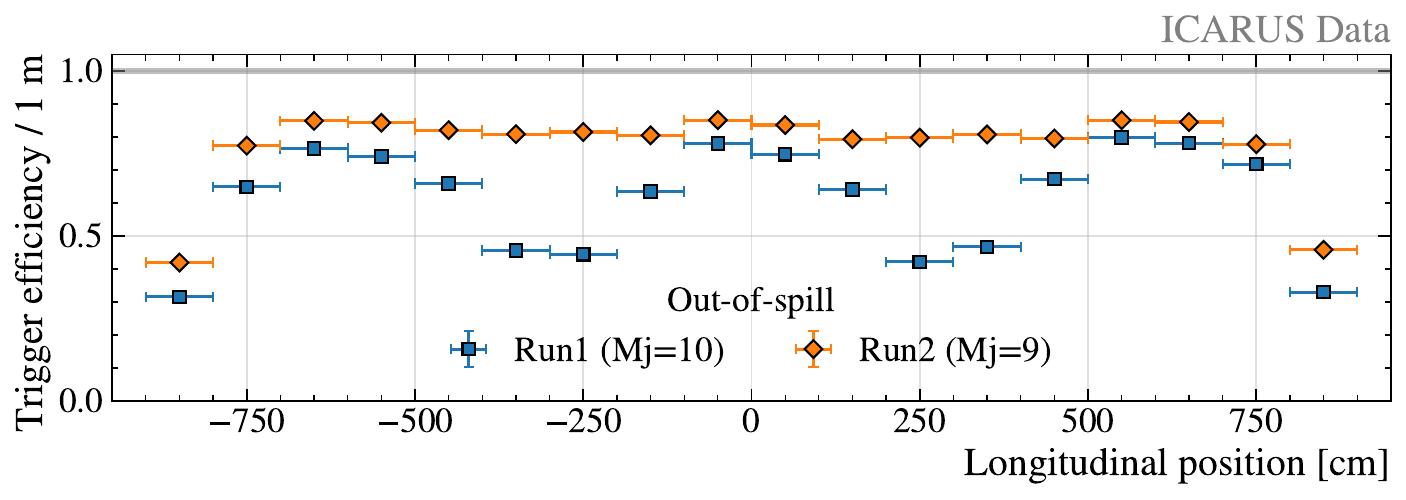}
    \caption{Trigger efficiency with \Mj{10} (Run1) and \Mj{9} (Run2) out-of-spills settings, measured with loosely-selected cosmic muons as function of the longitudinal average position of the tracks.
    The efficiency in \RunTwo{} with the five overlapped windows and \Mj{9} is significantly improved also in uniformity with respect to \RunOne{}, where only three tiled logic windows with \Mj{10} were used.}  
	\label{fig:efficiency-energy-run1-run2-outofspill}
\end{figure}

The corresponding trigger efficiency for collecting light from out-of-spill cosmic rays satisfying \Mj{10} (\Mj{9}) in \RunOne{} (\RunTwo{}) is shown in \cref{fig:efficiency-energy-run1-run2-outofspill}, using a muon sample selected loosely, independently of the particle orientation (see \cref{par:LooseSelection}).
Compared to \RunOne{}, the \RunTwo{} out-of-spill trigger efficiency is improved also in uniformity, by including the two additional overlapping windows into the trigger logic (\cref{sec:trig_logic}).

\section{Conclusions}

The ICARUS trigger is based on the detection of scintillation light produced by ionizing particles in liquid argon, in time coincidence with the proton extractions from BNB and NuMI beams at Fermilab.
The trigger system was deployed, tested and successfully put into operation for physics, taking data with the BNB and NuMI beams and cosmic rays with the detector fully commissioned in 2022.

The adopted trigger logic for the event recognition inside each module looks for the multiplicity of PMT-pair signals in \m{6} spatial windows, longitudinally spanning the \m{18} internal TPC length (PMT-majority trigger).
Minimum-bias triggers generated in correspondence of the beam gate, regardless of the scintillation light signals, are also collected routinely to provide unbiased data samples for the PMT trigger efficiency measurement and detector calibration.

The trigger efficiency was quantified in terms of the energy and position of collected cosmic rays.
The combined use of the external CRT cosmic tagging system with the PMT and TPC sub-detectors enabled the reconstruction and identification of clear cosmic ray interactions in the detector active volume.
Cosmic muons stopping in the liquid argon were selected in minimum-bias data to evaluate the trigger efficiency as a function of the reconstructed energy, in the same energy range as BNB neutrinos.

The trigger efficiency for a majority-5 of fired PMT pairs is found to be almost full for stopping muons above \MeV{300} in both modules, with remarkable spatial uniformity, and it reaches up to $>80 \%$ between 200 and \MeV{300}.
Analogously, the efficiency of the stricter, majority-9 trigger,
used in \RunTwo{} to identify cosmic rays crossing the detector during the \ms{1} TPC drift time, is found to be $>80\%$ for \MeV{>400} energy deposition.

\acknowledgments

This document was prepared by the ICARUS Collaboration using the resources of the Fermi National Accelerator Laboratory (Fermilab),
a U.S. Department of Energy, Office of Science, HEP User Facility.
Fermilab is managed by FermiForward Discovery Group, LLC, acting under Contract No. 89243024CSC000002.
This work was supported by the US Department of Energy, Istituto Nazionale di Fisica Nucleare (INFN, Italy), EU Horizon 2020 Research and Innovation Program under the Marie Sklodowska-Curie Grant Agreement Nos. 734303, 822185, 858199, and 101003460 and Horizon Europe Program research and innovation programme under the Marie Sklodowska-Curie Grant Agreement No. 101081478. In particular the fundamental support of INFN to ICARUS program, that allowed the development of the new LAr-TPC detection technique, is recognized and acknowledged. 
The ICARUS Collaboration would like to thank the MINOS Collaboration for having provided the Side CRT panels as well as Double Chooz (University of Chicago) for the Bottom CRT panels. The contribution of SBND colleagues, in particular for the development of a number of simulation, reconstruction and analysis tools which are shared within the SBN program, is warmly acknowledged. Finally, our experiment could not have been carried out without the support of CERN in the detector overhauling within the Neutrino Platform framework and of Fermilab in the detector installation and commissioning, and in providing the BNB and NuMI beams.

\bibliographystyle{JHEP}
\bibliography{refs}

\providecommand{\href}[2]{#2}\begingroup\raggedright\begin{thebibliography}{10}

\bibitem{acciarri2015proposaldetectorshortbaselineneutrino}
R.~Acciarri, C.~Adams, R.~An, C.~Andreopoulos, A.M.~Ankowski, M.~Antonello
  et~al., \emph{A proposal for a three detector short-baseline neutrino
  oscillation program in the {Fermilab} booster neutrino beam},
  \href{https://arxiv.org/abs/1503.01520}{{\ttfamily 1503.01520}}.

\bibitem{Abratenko2023}
P.~Abratenko, A.~Aduszkiewicz, F.~Akbar, M.A.~Pons, J.~Asaadi, M.~Aslin et~al.,
  \emph{{ICARUS at the Fermilab Short-Baseline Neutrino program: initial
  operation}}, \href{https://doi.org/10.1140/epjc/s10052-023-11610-y}{\emph{The
  European Physical Journal C} {\bfseries 83} (2023) 467}.

\bibitem{Aduszkiewicz_2025}
A.~Aduszkiewicz, L.~Bagby, B.~Behera, P.~Bernardini, S.~Bertolucci,
  M.~Betancourt et~al., \emph{{Design and implementation of the cosmic ray
  tagger system for the ICARUS detector at FNAL}},
  \href{https://doi.org/10.1088/1748-0221/20/04/T04002}{\emph{Journal of
  Instrumentation} {\bfseries 20} (2025) T04002}.

\bibitem{PMTS:Ali-Mohammadzadeh_2020}
B.~Ali-Mohammadzadeh, M.~Babicz, W.~Badgett, L.~Bagby, V.~Bellini, R.~Benocci
  et~al., \emph{{Design and implementation of the new scintillation light
  detection system of ICARUS T600}},
  \href{https://doi.org/10.1088/1748-0221/15/10/T10007}{\emph{Journal of
  Instrumentation} {\bfseries 15} (2020) T10007}.

\bibitem{icaruscollaboration2024searchhiddensectorscalar}
{\scshape ICARUS} collaboration, \emph{Search for a hidden sector scalar from
  kaon decay in the dimuon final state at {ICARUS}},
  \href{https://doi.org/10.1103/PhysRevLett.134.151801}{\emph{Phys. Rev. Lett.}
  {\bfseries 134} (2025) 151801}.

\bibitem{RWM}
M.~Backfish, \emph{{MiniBooNE Resistive Wall Current Monitor}},  Tech. Rep.
  \href{https://doi.org/10.2172/1128043}{TM-2556-AD}, Fermilab (2013).

\bibitem{Vicenzi:2025bvh}
{\scshape ICARUS} collaboration, \emph{{Calibration and timing performance of
  the light detection system in the ICARUS detector}},
  \href{https://doi.org/10.1088/1748-0221/20/03/C03049}{\emph{Journal of
  Instrumentation} {\bfseries 20} (2025) C03049}.

\bibitem{CAEN:a2795-manual}
CAEN, \emph{User manual {A2795} (liquid argon {TPC} readout board)},  October,
  2019.

\bibitem{miniboone_booster_description}
{\scshape MiniBooNE} collaboration, \emph{{Neutrino flux prediction at
  MiniBooNE}}, \href{https://doi.org/10.1103/PhysRevD.79.072002}{\emph{Phys.
  Rev. D} {\bfseries 79} (2009) 072002}.

\bibitem{WhiteRabbit}
J.~Serrano, M.~Cattin, E.~Gousiou, E.~van~der Bij, T.~Wlostowski, G.~Daniluk
  et~al., \emph{{The White Rabbit project}},  2013.

\bibitem{biery2018flexible}
K.~Biery, E.~Flumerfelt, J.~Freeman, W.~Ketchum, G.~Lukhanin, A.~Lyon et~al.,
  \emph{Flexible and scalable data-acquisition using the artdaq toolkit},
  \href{https://arxiv.org/abs/1806.07250}{{\ttfamily 1806.07250}}.

\bibitem{Marshall2015}
J.S.~Marshall and M.A.~Thomson, \emph{{The Pandora software development kit for
  pattern recognition}},
  \href{https://doi.org/10.1140/epjc/s10052-015-3659-3}{\emph{The European
  Physical Journal C} {\bfseries 75} (2015) 439}.

\bibitem{Acciarri2018}
R.~Acciarri, C.~Adams, R.~An, J.~Anthony, J.~Asaadi, M.~Auger et~al.,
  \emph{{The Pandora multi-algorithm approach to automated pattern recognition
  of cosmic-ray muon and neutrino events in the MicroBooNE detector}},
  \href{https://doi.org/10.1140/epjc/s10052-017-5481-6}{\emph{The European
  Physical Journal C} {\bfseries 78} (2018) 82}.

\bibitem{Abud2023}
A.A.~Abud, B.~Abi, R.~Acciarri, M.A.~Acero, M.R.~Adames, G.~Adamov et~al.,
  \emph{{Reconstruction of interactions in the ProtoDUNE-SP detector with
  Pandora}}, \href{https://doi.org/10.1140/epjc/s10052-023-11733-2}{\emph{The
  European Physical Journal C} {\bfseries 83} (2023) 618}.

\bibitem{ICARUS:2024calibration}
P.~Abratenko, N.~Abrego-Martinez, A.~Aduszkiewicz, F.~Akbar, L.~{Aliaga
  Soplin}, M.~{Artero Pons} et~al., \emph{{Calibration and simulation of
  ionization signal and electronics noise in the ICARUS liquid argon time
  projection chamber}},
  \href{https://doi.org/10.1088/1748-0221/20/01/P01032}{\emph{Journal of
  Instrumentation} {\bfseries 20} (2025) P01032}.

\bibitem{boone:2024fdy}
T.N.~Boone, \emph{{ICARUS cosmic ray tagger efficiency}}, Ph.D. thesis,
  Colorado State U., Fort Collins, 2024.

\bibitem{ICARUS:2024recombination}
P.~Abratenko, N.~Abrego-Martinez, A.~Aduszkiewicz, F.~Akbar, L.~Aliaga~Soplin,
  M.~Artero~Pons et~al., \emph{Angular dependent measurement of electron-ion
  recombination in liquid argon for ionization calorimetry in the {ICARUS}
  liquid argon time projection chamber},
  \href{https://doi.org/10.1088/1748-0221/20/01/P01033}{\emph{Journal of
  Instrumentation} {\bfseries 20} (2025) P01033}.

\end{thebibliography}\endgroup

\end{document}